\documentclass[conference]{IEEEtran}
\usepackage{amsmath,amssymb,amsfonts}
\usepackage[ruled,vlined,linesnumbered]{algorithm2e}
\usepackage{array}
\usepackage{tabularx}
\usepackage{booktabs}
\usepackage{cite}
\usepackage{graphicx}
\usepackage{subfigure}
\usepackage{xcolor}
\usepackage{url}
\usepackage{balance}
\graphicspath{{figures/}}
\begin{document}

\title{Continuum: Boundary-State-Aware Deterministic Scheduling in Dynamic LEO Networks}

\author{
\IEEEauthorblockN{
Siqi Yang\textsuperscript{1},
Zonghui Li\textsuperscript{2,*},
Chaoqun You\textsuperscript{1,*},
Yue Gao\textsuperscript{1}
}
\IEEEauthorblockA{
\textsuperscript{1} College of Computer Science and Artificial Intelligence,
Fudan University\\
Institute of Space Internet, Fudan University,
Shanghai 200438, China\\
Email: \{sqyang23, chaoqunyou, gao.yue\}@fudan.edu.cn\\
\textsuperscript{2} Beijing Key Lab of Traffic Data Analysis and Mining\\
School of Computer Science and Technology,
Beijing Jiaotong University, Beijing 100044, China\\
Email: lizonghui@bjtu.edu.cn
}
}

\maketitle

\begin{abstract}
Low Earth Orbit (LEO) constellations support mission-critical services requiring bounded latency and predictable delivery. However, conventional snapshot-based schedulers discard the forwarding progress of unfinished time-triggered traffic across topology transitions, degrading schedulability and timing consistency.
We present Continuum, a boundary-state-aware rolling-horizon scheduling framework. Continuum inherits buffered-node and arrival-time information, jointly optimizing remaining routes and transmission slots, and refines latency to improve timing consistency while preserving original timing constraints. 
Ablations show up to 16.6 percentage points higher
schedulability from boundary-state inheritance and a
10.9--37.7$\times$ reduction in mean jitter from joint
alignment. Scalability experiments evaluate synthesis performance for workloads of up to 10K flows.
\end{abstract}


\section{Introduction}
Low Earth Orbit (LEO) constellations are becoming an important part of space-air-ground integrated networks, supporting global sensing, remote control, and mission-critical
services~\cite{iridium,starlink,popovski2022perspective,tian2024large,hu2024time}. Such services require predictable packet delivery within end-to-end deadlines. Time-Sensitive Networking (TSN) and related deterministic networking techniques address these requirements through coordinated resource reservation and time-triggered (TT) scheduling ~\cite{kopetz2003time,finn2018introduction, steiner2010evaluation}.

TT scheduling determines routes and transmission slots before execution, enabling contention-free forwarding and predictable end-to-end latency. However, these schedules depend on the network conditions
under which they are generated. In LEO networks, satellite mobility changes inter-satellite connectivity and propagation delay~\cite{han2022time,ma2023network}, potentially invalidating a scheduled route before its traffic completes forwarding.

This creates a cross-snapshot continuity problem beyond ordinary route selection. A packet may have already reached an intermediate satellite when the topology changes, leaving valid forwarding progress but an infeasible remaining route. Restarting scheduling can discard this progress, whereas blindly continuing the previous schedule may violate link validity, deadline, or buffer-residence constraints. 

We present Continuum, a boundary-state-aware rolling-horizon scheduling framework. It inherits the buffered node and accumulated timing information of unfinished traffic,
refreshes the remaining route under the updated topology, and jointly evaluates path selection and transmission-slot allocation. Latency refinement further reduces cross-snapshot delay variation while retaining the original timing constraints.

The main contributions are:
\begin{itemize}
    \item \textbf{Model:} We formulate deterministic TT scheduling model across topology snapshots with explicit buffering, route-refresh, and cross-snapshot timing constraints.

    \item \textbf{Framework:} We propose Continuum, which combines boundary-state inheritance, joint path-slot evaluation, and latency refinement in a rolling-horizon framework.

    \item \textbf{Evaluation:} We evaluate Continuum on Iridium and Starlink through baseline comparisons, mechanism ablations, boundary-transition stress tests, small-instance
    SMT references, and large-scale experiments.
\end{itemize}

\section{Related Work}

\subsection{Deterministic Scheduling and Resource Allocation}

Offline TT scheduling establishes contention-free transmission by coordinating routes and time-slot reservations ~\cite{craciunas2016scheduling,oliver2018ieee}. Joint
routing/scheduling~\cite{schweissguth2017ilp,Atallah2020Routing}, scalable synthesis~\cite{li2020smsdac,long2025fastscheduler},fault-tolerant scheduling~\cite {ma2023dacfault}, and aggregation-cycle design~\cite{liu2024dacecqf} improve
feasibility, efficiency, or reliability. These approaches
address important scheduling objectives, but their generated
plans must still be reconciled with changes in network
connectivity during execution.

\subsection{Deterministic Scheduling in Dynamic LEO Networks}

LEO scheduling studies have investigated cyclic
forwarding~\cite{ma2023time}, time-varying routing and
multipath reliability~\cite{jiang2023spatio,peng2024fastts},
and learning-driven scheduling and resource allocation
~\cite{wang2025learningdac,yang2026crtpreprint}. These studies address dynamic routing, reliability, and
resource allocation, while time synchronization has also
been investigated~\cite{yang2026eaons}. Continuum instead focuses on
preserving unfinished forwarding state across topology
transitions.

\begingroup

\section{System Model and Problem Formulation}
\label{sec:system_model}
\label{sec:model}

\subsection{Network Model and Hierarchical Abstraction}

The time-varying graph $\mathcal{G}(t)=(\mathcal{V},\mathcal{E}(t))$ contains satellites, ground stations, and active links. Predictable orbital motion permits dividing the operation period $\mathcal{T}_{\mathrm{sys}}$ into snapshots $\mathcal{K}=\{1,\ldots,K\}$. Snapshot $k$ covers $[T_{\mathrm{start}}^k,T_{\mathrm{end}}^k)$, with duration $\Delta T_k=T_{\mathrm{end}}^k-T_{\mathrm{start}}^k$. Its graph is $\mathcal{G}_k=(\mathcal{V},\mathcal{E}_k)$: a link belongs to $\mathcal{E}_k$ if and only if it remains viable throughout that interval. Connectivity is fixed within a snapshot, but propagation varies; we use the conservative bound
      \begin{equation}
        \delta_{u,v}^k =
        \max_{t\in[T_{\text{start}}^k,\,T_{\text{end}}^k)}
        \frac{\mathrm{dist}_{u,v}(t)}{c}.
    \end{equation}
Here, $c$ is propagation speed. Intra-plane links
$\mathcal{E}_{\mathrm{intra}}$ connect adjacent satellites
within a plane with relatively stable geometry, while
inter-plane links $\mathcal{E}_{\mathrm{inter}}^k$ connect
adjacent planes and are more sensitive to delay variation
and handovers. Fig.~\ref{fig:continuity} shows this challenge:
a link scheduled in snapshot $k$ may be unavailable in
$k+1$, requiring unfinished traffic to resume at its
buffered node under the updated topology.

\begin{figure}[t]
    \centering
    \includegraphics[width=0.95\linewidth]{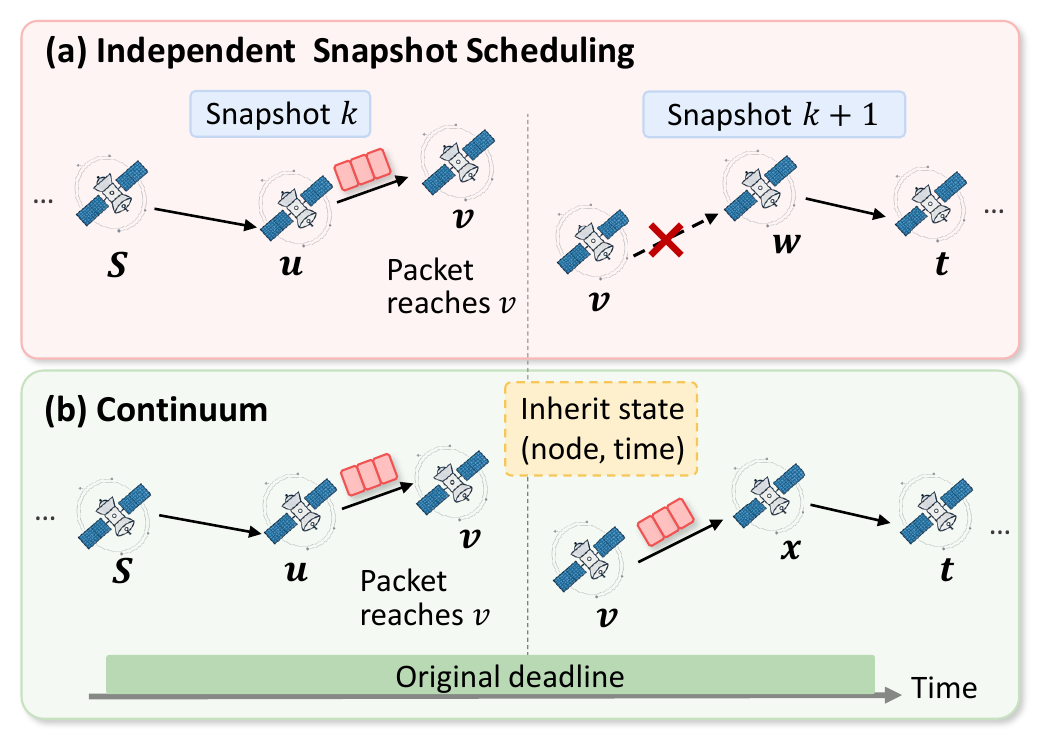}
    \caption{An example of cross-snapshot scheduling.}
    \label{fig:continuity}
\end{figure}

\subsection{Traffic Model}

Each critical TT flow $f_i\in\mathcal{F}$ is defined by
$f_i = \{s_i,d_i,P_i,C_i,D_i\}$, where $s_i,d_i$ are its endpoints,
$P_i$ its period, $C_i$ packet transmission time, computed as packet
length over link transmission rate, and $D_i$ its relative end-to-end deadline.

Let $n$ index a packet instance of flow $f_i$, with original
release time $T_{i,n}^{\mathrm{rel}}$ and release snapshot
$\kappa_{i,n}$. Its identity and release time remain unchanged
across snapshots. Let $\mathcal{N}_i$ denote the packet instances of flow
$f_i$ evaluated over the scheduling horizon, $\mathrm{loc}_{i,n}^{(k)}$ denote its
location at the start of snapshot $k$. Its remaining route is
selected from the feasible paths under the current topology:
\begin{equation}
    R_{i,n,k}\in
    \mathcal{R}_{i,k}
    \bigl(\mathrm{loc}_{i,n}^{(k)},d_i\bigr).
\end{equation}
If instance $(i,n)$ reaches a buffered node $v$ but cannot be
forwarded before snapshot $k$ ends, its location is retained
as $\mathrm{loc}_{i,n}^{(k+1)}=v$, and its remaining route is
refreshed under $\mathcal{G}_{k+1}$. 

\subsection{Problem Formulation}

For each active instance, the decisions are its route
$R_{i,n,k}$ and per-hop offsets $\phi_{i,k,n}^{(u,v)}$,
measured from the start of snapshot $k$.
Let $\pi_{\mathrm{proc}}$ be nodal processing time, $\pi_{\mathrm{guard}}$ the guard interval, $M_{\mathrm{buf}}$ the buffer residence-time bound, and $\mathcal{K}^{-}=\{1,\ldots,K-1\}$. Let $\mathcal{N}_{i,k}\subseteq\mathcal{N}_i$ denote the
active instances of flow $f_i$ in snapshot $k$. The
following constraints govern their scheduled hops,
including committed prefixes.

\subsubsection{Topology Validity Constraints}
The transmission of each hop must start and end within its snapshot:
\begin{equation}
\label{eq:noinflight}
\begin{aligned}
0\leq\phi_{i,k,n}^{(u,v)}
&\leq\Delta T_k-C_i-\pi_{\mathrm{guard}},\\
\phi_{i,k,n}^{(u,v)}+C_i+\delta_{u,v}^k
&\leq\Delta T_k,\quad (u,v)\in R_{i,n,k}.
\end{aligned}
\end{equation}

The second inequality explicitly disallows cross-snapshot in-flight propagation. Therefore, a transmission is executed only when the packet can safely reach the next hop within the same snapshot.

\subsubsection{Resource Contention Constraints}

All transmission instances sharing a directed link within
a snapshot must be serialized:

\begin{equation}
\begin{aligned}
&(u,v)\in R_{i,n,k}\cap R_{j,m,k}\\
&\Rightarrow\
\phi_{i,k,n}^{(u,v)}+C_i+\pi_{\mathrm{guard}}
\leq\phi_{j,k,m}^{(u,v)}\\
&\qquad\lor\
\phi_{j,k,m}^{(u,v)}+C_j+\pi_{\mathrm{guard}}
\leq\phi_{i,k,n}^{(u,v)}.
\end{aligned}
\end{equation}

\subsubsection{Path Dependency and Dynamic Propagation}
Consecutive scheduled hops obey store-and-forward causality:
\begin{equation}
\phi_{i,k,n}^{(v,w)}
\geq
\phi_{i,k,n}^{(u,v)}
+C_i+\delta_{u,v}^k+\pi_{\mathrm{proc}}.
\end{equation}
 
\subsubsection{Buffer Residence-Time Constraints}
Within a snapshot, the residence time at node $v$ between
consecutive hops is bounded by $M_{\mathrm{buf}}$:
\begin{equation}
\phi_{i,k,n}^{(v,w)}
-\left(\phi_{i,k,n}^{(u,v)}+C_i+\delta_{u,v}^k\right)
\leq M_{\mathrm{buf}}.
\end{equation}

\subsubsection{Cross-Snapshot Boundary Consistency}

Since cross-snapshot in-flight propagation is disallowed, an instance $(i,n)$ that cannot be further forwarded within snapshot $k$ is retained at its current node and deferred to snapshot $k+1$. Let $\mathcal{T}_{\mathrm{arr},i,n}^{(u,v),k}$ denote its snapshot-relative arrival offset at node $v$:
\begin{equation}
    \mathcal{T}_{\mathrm{arr},i,n}^{(u,v),k}
    =
    \phi_{i,k,n}^{(u,v)}+C_i+\delta_{u,v}^k.
\end{equation}
Its absolute arrival time is
$T_{i,n,\mathrm{arr}}^{(u,v),k}
=T_{\mathrm{start}}^k+\mathcal{T}_{\mathrm{arr},i,n}^{(u,v),k}$.

Let $B_{i,n}^{(k)}$ indicate buffering of instance $(i,n)$ at the end of snapshot $k$. If it reaches node $v$ within snapshot $k$ but cannot obtain a feasible next-hop transmission before the snapshot ends, it enters the buffered state:
\begin{equation}
\begin{split}
    &\mathcal{T}_{\mathrm{arr},i,n}^{(u,v),k}\leq\Delta T_k
    \ \land
    \mathrm{NoFeasibleNextHop}((i,n),v,k)\\
    &\Rightarrow B_{i,n}^{(k)}=1,\quad v\neq d_i.
\end{split}
\end{equation}

Here, $\mathrm{NoFeasibleNextHop}((i,n),v,k)$ is an operational predicate evaluated by the scheduler. It is true when no admissible next-hop transmission satisfies the current snapshot's topology, contention, residence-time, and deadline constraints.

Buffered instances retain their locations. For those
with feasible remaining routes, the selected route and
first-hop transmission satisfy:
\begin{equation}
\label{eq:boundary_inheritance}
\begin{split}
    B_{i,n}^{(k)}=1
    \ \Rightarrow\
    \mathrm{loc}_{i,n}^{(k+1)}=v,
    R_{i,n,k+1}\in\mathcal{R}_{i,k+1}(v,d_i),\\
    \pi_{\mathrm{proc}}
    \leq\phi_{i,k+1,n}^{(v,w)}
    \leq t_{\max},
    \hfill \forall k\in\mathcal{K}^{-}.
\end{split}
\end{equation}

where $(v,w)$ is the first hop of the selected remaining route, and
$t_{\max}=T_{i,n,\mathrm{buf}}^{(v)}
+M_{\mathrm{buf}}-T_{\mathrm{start}}^{k+1}$.
Upon buffering at node $v$, set
$T_{i,n,\mathrm{buf}}^{(v)}
=T_{i,n,\mathrm{arr}}^{(u,v),k}$ and retain it until
the instance leaves $v$. If no feasible remaining route exists, the instance is marked unschedulable.

\subsubsection{End-to-End Deadline Constraint}
For each packet instance $n$ of flow $f_i$, let
$T_{i,n}^{\mathrm{rel}}$ and $T_{i,n}^{\mathrm{fin}}$
denote its original absolute release time and final
completion time, respectively. Its end-to-end delay is:
\begin{equation}
    L_{i,n}
    =
    T_{i,n}^{\mathrm{fin}}-T_{i,n}^{\mathrm{rel}}.
\end{equation}
Let $y_{i,n}\in\{0,1\}$ indicate whether instance $(i,n)$
is selected for deadline-feasible completion. Define
$\mathcal{N}_i^{\mathrm{fin}}
=\{n\in\mathcal{N}_i:y_{i,n}=1\}$ and
$\mathcal{N}^{\mathrm{fin}}
=\{(i,n):f_i\in\mathcal{F},\
n\in\mathcal{N}_i^{\mathrm{fin}}\}$.
The deadline constraint is:
\begin{equation}
    L_{i,n}\leq D_i,\qquad
    \forall f_i\in\mathcal{F},\
    n\in\mathcal{N}_i^{\mathrm{fin}}.
\end{equation}

\subsubsection{Cross-Snapshot Latency Consistency Target}

For the same TT flow, the end-to-end delays of completed
instances released in different snapshots aare targeted to remain within the jitter tolerance $\eta_i$:
\begin{equation}
    |L_{i,n}-L_{i,m}|\leq\eta_i,
    \quad
    \forall f_i\in\mathcal{F},\
    n,m\in\mathcal{N}_i^{\mathrm{fin}},\
    \kappa_{i,n}\neq\kappa_{i,m}.
\end{equation}

This target captures latency consistency across
release snapshots rather than per-instance delay variation
during forwarding.

\subsection{Optimization Objective}
The optimization follows a lexicographic order: first maximize
the number of deadline-feasible TT flows, then minimize their
aggregate end-to-end latency. Let $\{\mathcal{F}_k\}$ partition
$\mathcal{F}$, where $\mathcal{F}_k$ contains flows first
released in snapshot $k$. Let $z_i\in\{0,1\}$ indicate whether
flow $f_i$ is schedulable over the evaluation horizon, enforced
by $z_i\leq y_{i,n}$ for all $n\in\mathcal{N}_i$.

\begin{equation}
    J_1^\star =
    \max
    \sum_{k\in\mathcal{K}}
    \sum_{f_i\in\mathcal{F}_k}
    z_i.
\end{equation}

Given the optimal schedulability value $J_1^\star$, the secondary objective minimizes the total latency of scheduled dataflows:
\begin{equation}
\begin{split}
    J_2^\star
    =\min\quad&
    \sum_{k\in\mathcal{K}}
    \sum_{f_i\in\mathcal{F}_k}
    z_i\sum_{n\in\mathcal{N}_i}L_{i,n},\\
    \text{s.t.}\quad&
    \sum_{k\in\mathcal{K}}
    \sum_{f_i\in\mathcal{F}_k}
    z_i=J_1^\star.
\end{split}
\end{equation}

This formulation prioritizes flow-level schedulability before latency minimization. A flow is counted only when all its evaluated packet instances meet their deadlines, and its instance-level latencies are included in the secondary objective only when the flow is scheduled.

\endgroup

\section{Continuum Scheduling Framework}
\label{sec:algorithm}

Continuum combines boundary-state inheritance, joint
path-slot evaluation using resource reservation tables (RRTs),
and latency refinement for cross-snapshot delay consistency. When a instance cannot complete within the current snapshot,
Continuum commits its longest feasible forwarding prefix
and carries the buffered node and absolute arrival time
to the next snapshot. The remaining route is then refreshed
under the updated topology, while the original deadline
and residence constraints remain in force.

\begin{algorithm}[t]
\footnotesize
\caption{Continuum Scheduling Algorithm}
\label{alg:continuum_main}
\SetKwInOut{Input}{Input}
\SetKwInOut{Output}{Output}
\Input{Snapshot sequence $\mathcal{K}$, flow set $\mathcal{F}$,
topology sequence $\{\mathcal{G}_k\}$}
\Output{Schedule $\Phi$ and completed-instance set $\mathcal{N}^{\mathrm{fin}}$}

$\Phi,\mathcal{N}^{\mathrm{fin}},\mathcal{L}_{\mathrm{next}}
\gets\emptyset$\;

\For{$k\in\mathcal{K}$}{
    Initialize $RRT[k]$ for active links\;
    $\mathcal{L}_{\mathrm{curr}}\gets\mathcal{L}_{\mathrm{next}}$;
    $\mathcal{L}_{\mathrm{next}}\gets\emptyset$\;
    Inherit buffered states; set $\mathrm{ready}_{i,n}^{(k)}$
    to $\pi_{\mathrm{proc}}$ (inherited) or
    $T_{i,n}^{\mathrm{rel}}-T_{\mathrm{start}}^k$ (new)\;
    $\mathcal{Q}_k\gets
    \mathcal{L}_{\mathrm{curr}}\cup\mathcal{N}_k^{\mathrm{new}}$\;
    Sort inherited instances first, then by descending priority score\;

    \For{each non-rejected $(i,n)\in\mathcal{Q}_k$}{
        Refresh routes and generate $\mathcal{C}_{i,n,k}$\;
        $best\gets\texttt{NULL}$; $c_{\min}\gets+\infty$\;

        \For{$R\in\mathcal{C}_{i,n,k}$}{
            $(sol,cost)\gets
            \textsc{TrialScheduleOnPath}
            ((i,n),R,k,RRT[k],\texttt{full})$\;
            \If{$sol\neq\texttt{NULL}$ and $cost<c_{\min}$}{
                $best\gets sol$; $c_{\min}\gets cost$\;
            }
        }

        \eIf{$best\neq\texttt{NULL}$}{
            Commit $best$ to $RRT[k]$ and $\Phi$\;
            Add $(i,n)$ to $\mathcal{N}^{\mathrm{fin}}$\;
        }{
            $(p,v_p,a_p)\gets
            \textsc{BestPrefix}((i,n),\mathcal{C}_{i,n,k},k,RRT[k])$\;
            \eIf{$p\neq\texttt{NULL}$ and
            $\textsc{CanDefer}((i,n),p,k)$}{
                Commit $p$ to $RRT[k]$ and $\Phi$\;
                Retain $(v_p,a_p)$ for instance $(i,n)$\;
                Mark $(i,n)$ as buffered and add it to
                $\mathcal{L}_{\mathrm{next}}$\;
            }{
                Mark $(i,n)$ as rejected\;
            }
        }
    }
}
Compute end-to-end latencies using original absolute release times\;
Perform window-level refinement on $\mathcal{N}^{\mathrm{fin}}$\;
\KwRet{$\Phi,\mathcal{N}^{\mathrm{fin}}$}\;
\end{algorithm}

\subsection{Boundary-State Inheritance and Rolling-Horizon Scheduling}

\subsubsection{Phase 1: Buffered-State Inheritance and Route Refresh}

At snapshot $k$, Algorithm~\ref{alg:continuum_main}
(lines~4--5) inherits buffered nodes, absolute arrival
times, and original timing constraints, while setting
local ready times to $\pi_{\mathrm{proc}}$. Routes are
refreshed from inherited nodes under updated topology
(line~9), and Algorithm~\ref{alg:trial_schedule}
(line~1) initializes each trial from its location and
ready time, using its retained buffer-arrival
time when available.

\subsubsection{Phase 2: Queue Construction and Urgency-Based Prioritization}

Algorithm~\ref{alg:continuum_main} (lines~6--7) prioritizes
inherited instances, then orders instances by descending priority
score. The feasibility margin and priority score are
\begin{equation}
    u_{i,n}^{(k)}
    =D_i-\left(\widehat{A}_{i,n}^{(k)}
    +\widehat{R}_{i,n}^{(k)}\right),
\end{equation}

where $\widehat{A}_{i,n}^{(k)}$ is the estimated accumulated delay and
$\widehat{R}_{i,n}^{(k)}$ is the minimum estimated residual delay to the destination. The priority score is

\begin{equation}
    Score_{i,n}^{(k)}
    =\frac{W_i}
    {\max\{u_{i,n}^{(k)},\epsilon\}}.
\end{equation}

where $W_i$ is the flow priority weight and $\epsilon>0$ avoids
division by zero. Smaller positive margins yield higher scores.

\subsubsection{Phase 3: Candidate-Path Generation and Joint Path-Slot Evaluation}

Algorithm~\ref{alg:continuum_main} (lines~9--14)
evaluates candidate paths using Algorithm~\ref{alg:trial_schedule}
(lines~2--22), which checks hop feasibility aand searches
the RRT for available transmission slots. The minimum-cost feasible assignment is selected:

\begin{equation}
\label{eq:pathcost}
\begin{split}
    \mathrm{Cost}_{i,n,k}^{(q)}
    ={}&\alpha\widehat{L}_{i,n,k}^{(q)}
      +\beta\widehat{W}_{i,n,k}^{(q)}
      +\gamma N_{\mathrm{inter}}^{(q)}\\
     &+\zeta\mathrm{SwitchPenalty}_{i,n,k}^{(q)}.
\end{split}
\end{equation}
where $\widehat{L}_{i,n,k}^{(q)}$ and $\widehat{W}_{i,n,k}^{(q)}$ denote predicted end-to-end latency and accumulated waiting cost, respectively; $N_{\mathrm{inter}}^{(q)}$ counts inter-plane links, and $\mathrm{SwitchPenalty}_{i,n,k}^{(q)}$ penalizes unnecessary route changes. The weights $\alpha,\beta,\gamma,\zeta$ rank feasible candidates.

\begin{algorithm}[t]
\footnotesize
\caption{TrialScheduleOnPath: Tentative Scheduling on a Candidate Path}
\label{alg:trial_schedule}
\SetKwInOut{Input}{Input}
\SetKwInOut{Output}{Output}
\Input{instance $(i,n)$, path $R_{i,n,k}^{(q)}$, snapshot $k$,
RRT $RRT[k]$, mode $m$}
\Output{Complete schedule and cost, or safe prefix with
terminal node and arrival time, otherwise \texttt{NULL}}

$v\gets\mathrm{loc}_{i,n}^{(k)}$; $t_{\min}\gets\mathrm{ready}_{i,n}^{(k)}$; $sol\gets\emptyset$; $a\gets T_{i,n,\mathrm{buf}}^{(v)}$ if buffered\;

\While{$v\neq d_i$}{
    $e\gets\textsc{GetNextLink}(R_{i,n,k}^{(q)},v)$;
    $Slot\gets\texttt{NULL}$\;

    \If{$e\neq\texttt{NULL}$ and $e\in\mathcal{E}_k$}{
        \eIf{$v=s_i$}{
            $t_{\max}\gets\Delta T_k-C_i-\pi_{\mathrm{guard}}$\;
        }{
            $t_{\max}\gets a+M_{\mathrm{buf}}-T_{\mathrm{start}}^k$\;
        }
        $Slot\gets\textsc{TreeQuery}(RRT[k],sol,e,C_i,t_{\min},t_{\max})$\;
    }

    \If{$Slot\neq\texttt{NULL}$}{
        $t_{\mathrm{arr}}\gets Slot.\mathrm{End}+\delta_e^k$\;
        \If{$t_{\mathrm{arr}}>\Delta T_k$ or
        $\textsc{MissDeadline}((i,n),e,t_{\mathrm{arr}},k)$}{
            $Slot\gets\texttt{NULL}$\;
        }
    }

    \If{$Slot=\texttt{NULL}$}{
        \eIf{$m=\texttt{full}$}{
            \KwRet{\texttt{NULL}}\;
        }{
            \textbf{break}\;
        }
    }

    Append $(e,Slot)$ to $sol$\;
    $a\gets T_{\mathrm{start}}^k+t_{\mathrm{arr}}$\;
    $t_{\min}\gets t_{\mathrm{arr}}+\pi_{\mathrm{proc}}$\;
    $v\gets\mathrm{Dst}(e)$\;
}

\eIf{$v=d_i$}{
    $cost\gets\textsc{PathCost}((i,n),R_{i,n,k}^{(q)},sol,k)$\;
    \KwRet{$(sol,cost)$}\;
}{
    \eIf{$m=\texttt{prefix}$ and $sol\neq\emptyset$}{
        \KwRet{$(sol,v,a)$}\;
    }{
        \KwRet{\texttt{NULL}}\;
    }
}
\end{algorithm}

\subsubsection{Phase 4: Buffered Stitching and Completion Collection}

If no complete candidate is feasible, Algorithm~\ref{alg:continuum_main}
invokes $\textsc{BestPrefix}$ and $\textsc{CanDefer}$
(lines~19--20). $\textsc{BestPrefix}$ uses
Algorithm~\ref{alg:trial_schedule} in prefix mode to select
the longest safe prefix, breaking ties by path cost.
Algorithm~\ref{alg:continuum_main} commits accepted prefixes
for the next snapshot (lines~21--23) and rejects others
(lines~24--25).

\subsection{Latency Refinement}
To improve end-to-end latency consistency across periodic
rounds, Continuum uses the maximum observed latency among
the completed instances $\mathcal{W}_i\subseteq
\mathcal{N}_i^{\mathrm{fin}}$ of flow $f_i$ in the scheduling
window as a reference and introduces feasible waiting to
align shorter rounds:

\begin{equation}
\label{equ1}
    L_i^{\mathrm{ref}}
    =
    \max_{n\in\mathcal{W}_i}L_{i,n}.
\end{equation}
The alignment gap for instance $n$ is
\begin{equation}
    \Delta_{i,n}
    =L_i^{\mathrm{ref}}-L_{i,n}.
\end{equation}

When $\Delta_{i,n}>0$, the safe waiting budget at hop $h$ is
\begin{equation}
    S_{i,n,h}
    =\min\left\{
       U_{i,n,h},
       M_{\mathrm{buf}}-\tau_{i,n,h},
       \Gamma_{i,n,h}
    \right\}.
\end{equation}
where
\begin{equation}
    U_{i,n,h}
    =D_i-\left(A_{i,n,h}+R_{i,n,h}\right).
\end{equation}
is the deadline-derived waiting slack.
Here, $\tau_{i,n,h}$ is the residence time already reserved
at node $h$, including time accumulated before snapshot
transitions, and $\Gamma_{i,n,h}$ is the available no-conflict slack.
The aggregate waiting budget for instance $n$ is
\begin{equation}
    S_{i,n}=\sum_h S_{i,n,h}.
\end{equation}

Alignment requires sufficient aggregate slack and a
jointly feasible allocation across hops. When the
aggregate budget is insufficient, the budget shortfall is
\begin{equation}
\label{equ2}
    \xi_{i,n}
    =\Delta_{i,n}-S_{i,n},
    \qquad S_{i,n}<\Delta_{i,n}.
\end{equation}
Per-hop budgets are coupled by deadline, contention,
residence, and no-in-flight constraints. If no feasible
allocation achieves the target, the remaining latency
mismatch is retained.

\begin{figure*}[!t]
\centering
\includegraphics[width=\textwidth]{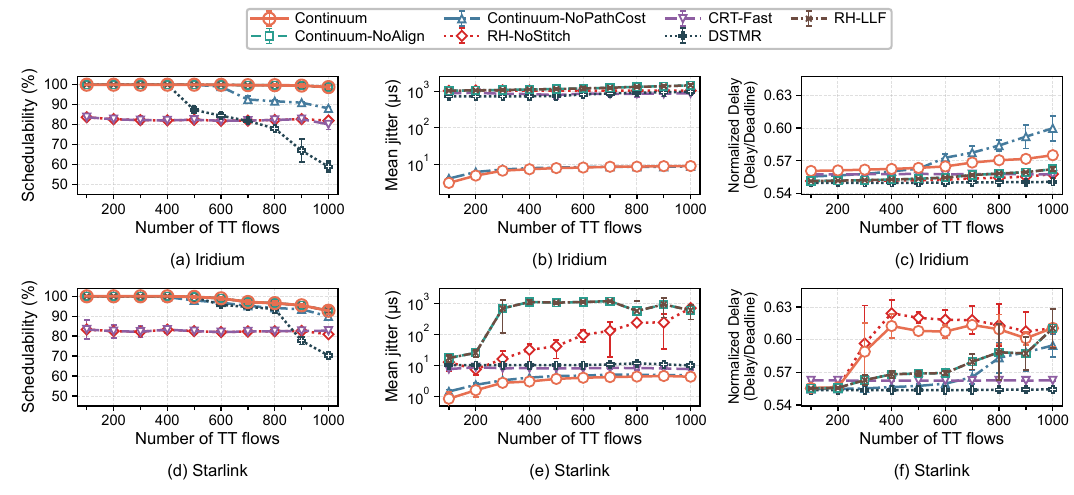}
\caption{Overall performance on Iridium and
Starlink: (a,d) schedulability, (b,e) mean jitter,
and (c,f) normalized end-to-end delay.}
\label{fig:overall}
\end{figure*}

\begin{figure*}[!t]
\centering
\setlength{\subfigtopskip}{0pt}
\setlength{\subfigbottomskip}{1pt}
\subfigure[Schedulability]{\includegraphics[width=0.48\textwidth,trim=0 3bp 0 2bp,clip]{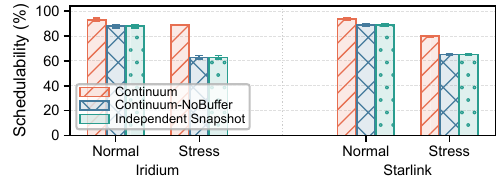}}\hfill
\subfigure[Mean jitter]{\includegraphics[width=0.48\textwidth,trim=0 3bp 0 2bp,clip]{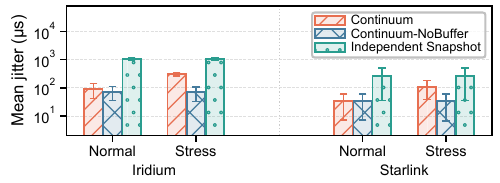}}
\caption{Boundary-transition stress test under normal and stress workloads (5\% and 30\% target boundary-crossing demand, respectively).}
\label{fig:transition_stress}
\end{figure*}

\subsection{Complexity Analysis}

Let $K$ be the number of snapshots, $A$ the maximum
number of active instances per snapshot, $H$ the maximum
candidate-path length, $Q$ the maximum number of candidate
paths per instance, and $S$ the maximum number of reserved
intervals on a link. Queue sorting costs
$O(|\mathcal{Q}_k|\log|\mathcal{Q}_k|)$ per snapshot.
Interval insertion costs $O(\log S)$, while earliest-slot
search is conservatively bounded by $O(S)$. Evaluating up
to $Q$ candidate paths per instance therefore costs
$O(KAQHS)$, and latency refinement adds at most $O(KAH)$.
The overall reservation-scheduling complexity is
\begin{equation}
    O\left(
    \sum_{k=1}^{K}|\mathcal{Q}_k|\log|\mathcal{Q}_k|
    +KAQHS+KAH
    \right).
\end{equation}
This bound covers scheduling over bounded candidate sets; the
total synthesis cost also includes candidate-path generation.
$Q=K_{\mathrm{cand}}$ controls the exploration overhead, without
implying that this candidate budget is optimal.

\section{Evaluation}
\label{sec:evaluation}
\subsection{Experimental Setup}

\subsubsection{Simulation Environment} The simulations were conducted on the Iridium and Starlink constellations. The Iridium constellation is a representative polar-orbit satellite network consisting of 6 orbit planes with 11 satellites per plane. The Starlink constellation consists of 1584 satellites in 72 orbits at 550~km~\cite{starlink}. The dynamic network topology is discretized into a sequence of 10 time slots, with each slot duration set to 10 seconds. To reflect a realistic multi-service network scenario, we assume the deterministic traffic operates within a dedicated network slice. Accordingly, the ISLs are configured with a bandwidth of 200 Mbps reserved specifically for TT flows.

\subsubsection{Traffic Generation and Candidate Paths} TT flows are generated by randomly selecting source-destination pairs from the satellite nodes. Periods are randomly selected from \{16ms, 32ms, 64ms, 128ms, 256ms, 512ms\}; Packet lengths are randomly assigned values between 64 and 1518 bytes. 
Candidate paths are generated using Yen’s K-shortest simple paths algorithm, with $K=5$. The four terms in Eq.~\eqref{eq:pathcost} use fixed weights $\alpha=1$, $\beta=0.2$, $\gamma=50$, and $\zeta=500$ across the corresponding Continuum experiments. Timing costs are measured in microseconds, and the route-switch penalty is a unit indicator. 
The main workload contains 18\% boundary-probe groups.
Periodic groups use periodic reservations, while boundary
probes generate one-shot instances. Their relative deadlines
are set to 1.8 and 2.8 times the maximum snapshot-wise
shortest-path total delay, respectively.

\subsubsection{Methods Compared}

We compare Continuum with representative scheduling baselines
and mechanism ablations.
\textbf{CRT-Fast}~\cite{yang2026crtpreprint} uses conflict-aware
path selection and worst-case-delay-based admission.
\textbf{DSTMR}~\cite{jiang2023spatio} combines spatio-temporal
routing, redundant coding, and multipath scheduling.
\textbf{RH-LLF} prioritizes inherited flows using unweighted
least-laxity-first ordering, without latency alignment.
\textbf{RH-NoStitch} disables state inheritance and alignment,
rejecting unfinished rounds within their release snapshot.
\textbf{Continuum-NoPathCost} accepts the first feasible complete
candidate while retaining prefix selection.
\textbf{Continuum-NoAlign} disables alignment-aware ranking and
latency refinement. \textbf{Continuum-NoBuffer} disables state
inheritance but retains path selection and alignment.
\textbf{Continuum-NoJoint} selects the minimum-delay path before
slot allocation, without alternative-path fallback.
\textbf{Independent Snapshot} performs joint path-slot evaluation
within each snapshot but inherits neither unfinished forwarding
state nor historical path/latency references.

\subsection{Overall Performance of Continuum}

At 1000 groups, Continuum, Continuum-NoAlign, and RH-LLF
achieve identical schedulability: $98.80\%$ on Iridium and
$92.77\%$ on Starlink. Continuum exceeds RH-NoStitch by
$16.83$ and $11.80$ percentage points, respectively
(Fig.~\ref{fig:overall}(a,d)).

Continuum achieves mean jitter of $8.91$~$\mu$s on Iridium
and $4.34$~$\mu$s on Starlink, compared with $1454.84$~$\mu$s
and $588.42$~$\mu$s for Continuum-NoAlign
(Fig.~\ref{fig:overall}(b,e)). Continuum-NoPathCost achieves slightly lower
jitter on Iridium ($8.74$~$\mu$s), but with lower schedulability, illustrating the trade-off between completion
and timing consistency.

Fig.~\ref{fig:overall}(c,f) compares the normalized end-to-end delay of scheduled flows. Continuum's latency refinement introduces
waiting to reduce delay variation, trading some latency for improved timing consistency relative to Continuum-NoAlign.

\subsection{Mechanism Ablation}

In a separate three-seed ablation campaign at 1000 TT flows,
removing boundary-state inheritance reduces schedulability
by $16.60$ percentage points on Iridium and $14.97$ on
Starlink (Table~\ref{tab:mechanisms}).
Independent Snapshot matches Continuum-NoBuffer in schedulability,
as both reject unfinished rounds rather than inherit their
forwarding progress. Joint path-slot evaluation improves
schedulability over NoJoint by $19.5$ and $17.8$
percentage points, respectively, showing the benefit of
considering slot feasibility during route selection.
Continuum-NoAlign retains Continuum's schedulability rates
but increases mean jitter by $37.7\times$ on Iridium and
$10.9\times$ on Starlink. This indicates that alignment-aware
ranking and latency refinement jointly improve timing
consistency, while boundary-state inheritance and joint
path-slot evaluation contribute to schedulability.

\begin{table}[!t]
\centering\footnotesize
\caption{Boundary-state continuity ablation at 1000 TT flows. Values show mean $\pm$ SD over three seeds.}
\label{tab:mechanisms}
\setlength{\tabcolsep}{4pt}
\begin{tabular}{@{}lrr@{}}
\toprule
Method & schedulability (\%) & Jitter ($\mu$s) \\
\midrule
\multicolumn{3}{l}{\textit{Iridium}} \\
\textbf{Continuum} & $\boldsymbol{87.67\pm1.16}$ & $26.46\pm8.00$ \\
Continuum-NoBuffer & $71.07\pm0.81$ & $\boldsymbol{24.62\pm7.52}$ \\
Continuum-NoJoint & $68.13\pm0.64$ & $528.79\pm133.39$ \\
Continuum-NoAlign & $\boldsymbol{87.67\pm1.16}$ & $997.56\pm43.29$ \\
Independent snapshot & $71.07\pm0.81$ & $773.98\pm42.39$ \\
\midrule
\multicolumn{3}{l}{\textit{Starlink}} \\
\textbf{Continuum} & $\boldsymbol{86.37\pm0.91}$ & $99.87\pm39.40$ \\
Continuum-NoBuffer & $71.40\pm1.06$ & $20.58\pm10.76$ \\
Continuum-NoJoint & $68.57\pm0.47$ & $\boldsymbol{11.20\pm8.12}$ \\
Continuum-NoAlign & $\boldsymbol{86.37\pm0.91}$ & $1090.77\pm100.90$ \\
Independent snapshot & $71.40\pm1.06$ & $220.45\pm44.33$ \\
\bottomrule
\end{tabular}
\end{table}

\subsection{Boundary-Transition Stress Test}
The test uses 700 periodic groups and 300 boundary-probe groups, each generating five one-shot instances. Normal and Stress place 50 and 300 probes near snapshot boundaries, respectively (5\% and 30\% of all groups), only probe release times differ. Under Stress, the actual crossing rates are $25.2\%$--$27.7\%$ on Iridium and $13.4\%$--$14.6\%$ on Starlink. Under Normal, they are $5.0\%$ and $4.3\%$--$5.0\%$, respectively. Both non-inheriting baselines record zero crossings.

Under Stress, Fig.~\ref{fig:transition_stress}(a) shows that Continuum achieves schedulability of $88.70\%$ on Iridium and $79.77\%$ on Starlink, versus $62.80\%$ and $65.20\%$ for both non-inheriting baselines, respectively. Fig.\ref{fig:transition_stress}(b) compares mean group jitter over the common completed groups across methods and workloads.

\subsection{SMT Validation and Scalability}

\begin{figure}[!t]
\centering
\includegraphics[width=0.85\linewidth]{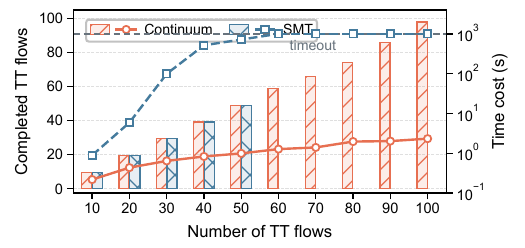}
\caption{Performance comparison between Continuum and SMT.}
\label{fig:exact}
\end{figure}

\begin{figure}[!t]
\centering
\includegraphics[width=0.85\linewidth,trim=0 11bp 0 0,clip]{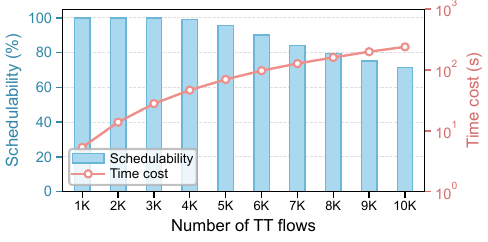}
\caption{Scalability performance of the Continuum.}
\label{fig:scale}
\end{figure}

Continuum matches SMT admission counts on all solved
instances up to 50 flows (Fig.~\ref{fig:exact}).
As the flow count increases, SMT synthesis becomes
substantially more expensive and reaches the 1000-s
per-stage timeout at 60 flows, whereas Continuum
schedules 59 flows in 1.27~s. This demonstrates the
computational advantage of Continuum over SMT.
We stress-test Continuum on Starlink with up to 10K (Fig.~\ref{fig:scale}). Continuum
maintains strong schedulability over a wide range of largescale traffic loads, achieving $100\%$ at 1K flows and $71.37\%$ at 10K flows. This result demonstrates
that Continuum remains effective in dense megaconstellation scenarios.

\section{Conclusion}
This paper presents Continuum, a boundary-state-aware
rolling-horizon scheduling framework for deterministic TT
communication in dynamic LEO networks. By inheriting unfinished
forwarding state, jointly evaluating routes and transmission
slots, and refining latency, Continuum preserves scheduling
continuity across topology transitions. Experiments on Iridium
and Starlink demonstrate the effectiveness of the proposed
framework in improving schedulability and timing consistency
under dynamic topology.

\clearpage
\balance

\bibliographystyle{IEEEtran}
\bibliography{ref,dac_references}

\end{document}